# Averaging of Equations for Flow and Transport in Random Porous Media

Mark Shvidler and Kenzi Karasaki

Earth and Environmental Sciences Area, Lawrence Berkeley National Laboratory, Berkeley, CA 94720; Mshvidler@lbl.gov, kkaraski@lbl.gov

**Abstract**

In the papers (Shvidler, 1985, 1993; and Shvidler and Karasaki, 1999, 2001, 2005, and 2008) we developed an approach for finding the exactly averaged equations of flow and transport in porous media. We studied for steady state flow with sources and also analyzed unsteady flow and non-reactive solute transport in unbounded domains with stochastically homogeneous conductivity random fields.

In the base of the approach is the existence of appropriate random Green's functions and relevant linear random operators. Examination of random fields with global symmetry makes it possible to describe some different types of averaged equations with non-local unique vector or tensor operators.

In this paper we substantially extend the approach and study more general processes in bounded or unbounded fields of any dimensions and examine the cases where the random fields of conductivity and porosity are stochastically homogeneous or non-homogeneous.

## 1. Introduction

The operator approach for analyzing some fields, and in particular the problems of flow in porous media, has been applied in certain studies (Finkelberg, 1964; Shermergor, 1977; Shvidler, 1985, 1993; Neuman and Orr, 1993; Zhikov et al., 1994; Tartakovsky and Neuman, 1998; Indelman, 2002, etc). As we proved in Shvidler and Karasaki (1999, 2001, 2005, and 2008) the exact equation for flow-velocity in the case of a stochastically homogeneous unbounded field contains the convolution of a unique kernel-vector and the head-function, or the convolution reduced to a unique and appropriate symmetric tensor and the vector of head gradient.

As in our previous publications here we used appropriate Green's functions. We studied steady flow, unsteady flow and transport of solute in unbounded space, as well as steady flow and transient flow in bounded space. For determination of the averaged flow velocity we use the operational relations between local random head and the global field of averaged head. For finding the averaged solute flux, we use the operational relations between random concentration of solute and the global field of averaged concentration of solute. In this way we find the exact averaged equations without directly solving the stochastic equations.

## 2. Steady Flow

We consider a steady flow with sources and sinks that are distributed in a porous, single-connected heterogeneous, bounded or unbounded d-dimensional domain $\Omega$. The local condition of flow continuity and Darcy's law, the boundary condition are given by the following equations:

$$\nabla \mathbf{v}(\mathbf{x}) = f(\mathbf{x}), \quad \mathbf{v}(\mathbf{x}) = -\boldsymbol{\sigma}(\mathbf{x})\nabla u(\mathbf{x}), \quad \mathbf{x} \in \Omega \tag{2.1}$$

$$u(\mathbf{x})=0, \quad \mathbf{x}\in\partial\Omega \tag{2.2}$$

Here, $\mathbf{x}$ is a vector, $u(\mathbf{x})$ is the reduced pressure or hydraulic head, $\mathbf{v}(\mathbf{x})$ is the Darcy's velocity vector, the random second rank conductivity tensor $\boldsymbol{\sigma}(\mathbf{x})$ is symmetric by subscript. The function $f(\mathbf{x})$ is the given non-random density of flow sources and sinks, which is an integrable and compactly supported function or a distribution with bounded support.

Our target is to find the relationship between the mean fields $U(\mathbf{x})=\langle u(\mathbf{x})\rangle$ and $\mathbf{V}(\mathbf{x})=\langle \mathbf{v}(\mathbf{x})\rangle$.
Combining now Equations (2.1) and (2.2) we have the elliptic equation and boundary condition for the scalar function $u(\mathbf{x})$

$$-\nabla\big[\boldsymbol{\sigma}(\mathbf{x})\nabla u(\mathbf{x})\big]=f(\mathbf{x}), \mathbf{x}\in\Omega, \quad u(\mathbf{x})=0, \mathbf{x}\in\partial\Omega \tag{2.3}$$

Let $u(\mathbf{x})$-the random solution of (2.3) exist and be unique. For this solution we can write:

$$lu = f \tag{2.4}$$

where the random operator $l=-\nabla\boldsymbol{\sigma}(\mathbf{x})\nabla$ maps the random spaces of functions $u(\mathbf{x})$ onto the nonrandom space of functions $f(\mathbf{x})$.

In accordance with the Green's formula we have the unique solution of system (2.3)

$$u(\mathbf{x})=\int_\Omega g(\mathbf{x},\mathbf{y})f(\mathbf{y})dy^d \tag{2.5}$$

where the kernel $g(\mathbf{x},\mathbf{y})$ is a random Green's function that satisfies the equations

$$lg(\mathbf{x},\mathbf{y})=\delta(\mathbf{x}-\mathbf{y}), \quad \mathbf{x},\mathbf{y}\in\Omega; \quad g(\mathbf{x},\mathbf{y})=0, \mathbf{x}\in\partial\Omega, \mathbf{y}\in\Omega \tag{2.6}$$

Now we write the unique solution of Equations (2.3) and (2.4) in a symbolic form:

$$u=mf, \quad m=\int_\Omega g(\mathbf{x},\mathbf{y})(\bullet)dy^d \tag{2.7}$$

(The symbol ($\bullet$) denotes that the operator acts on respective entity.)

Applying the random operator $l$ to the equation (2.7) and comparing the result with Equation (2.4), we have $lm=\mathrm{I}^{(f)}$. Here $\mathrm{I}^{(f)}$ is the identity mapping of the space of the local functions $f$. Applying the operator $m$ to Equation (2.4) and compare the result with Equation (2.7), we have $ml=\mathrm{I}^{(u)}$. Here $\mathrm{I}^{(u)}$ is the identity mapping for the space of the random local functions $u$. Thus the random operators $l$ and $m$ are non-commutative because their products-the identity operators $\mathrm{I}^{(f)}$ and $\mathrm{I}^{(u')}$ are defined in different functional spaces.

So, averaging Equation (2.7) over the ensemble of $\boldsymbol{\sigma}(\mathbf{x})$, we have for fields $U=\langle u\rangle$ and $G=\langle g\rangle$

$$U=\langle m\rangle f, \quad \langle m\rangle=\int_\Omega G(\mathbf{x},\mathbf{y})(\bullet)dy^d, \quad G(\mathbf{x},\mathbf{y})=\langle g(\mathbf{x},\mathbf{y})\rangle \tag{2.8}$$

Since averaging over the probability is related to summation, the scalar operator $\langle m\rangle$ exists and is nonsingular, the unique scalar operator $\langle m\rangle^{-1}=\mathbf{L}$ exists and, when applied to Equation (2.8), we have for $U$ the averaged equation and the boundary condition

$$\mathbf{L}U = f, \quad \mathbf{x} \in \Omega \quad \text{and} \quad U_{\partial\Omega} = 0 \tag{2.9}$$

The so-called effective nonrandom scalar operator $\mathbf{L}$ in the general case is nonlocal and unique.

Obviously operators $\langle m \rangle$ and $\mathbf{L} = \langle m \rangle^{-1}$ are non-commutative because $\langle m \rangle \mathbf{L} = \mathrm{I}^{(U)}$ and $\mathbf{L}\langle m \rangle = \mathrm{I}^{(f)}$. Comparing Equations (2.7) and (2.8) we obtain the relationship between the local random field $u$ and the nonrandom field $U$

$$u = nU \tag{2.10}$$

Here the nonlocal random operator $n$ is the random nonlocal operator $m$ which is normalized on the average by the nonlocal and nonrandom operator $\mathbf{L}^{-1} = \langle m \rangle$

$$n = m\langle m \rangle^{-1} \tag{2.11}$$

It is clear that $n$ is a random operator that map the space of non-random functions $U$ into the space of the random functions $u$. In contrast the random double-point operator $n^{-1} = \langle m \rangle m^{-1}$ maps almost all the spaces of relevant realizations of random function $u$ into the space of the nonrandom function $U$. It is evident from equation (2.10) that the mean operator $\langle n \rangle = \mathrm{I}^U$ is the identity mapping of the functions $U$.

Using the rule of operator multiplication, and bearing in mind that the double point operator $n$ depends on the variable $\mathbf{x}$ and conversely, except for the case of homogenization limit, that the scalar field $U$ in Equation (2.10) is independent of $\mathbf{x}$, we have from equation $(2.1)$

$$\mathbf{v} = \boldsymbol{\pi}U, \quad \boldsymbol{\pi} = -\boldsymbol{\sigma}\nabla n \tag{2.12}$$

The composition of the random operators $-\boldsymbol{\sigma}\nabla$ and $n$ maps the space of the nonrandom functions $U$ in the space of the random vectors $\mathbf{v}$.

Averaging now Equations (2.12), we derive the mean nonrandom velocity vectors

$$\mathbf{V} = \boldsymbol{\Pi}U, \quad \boldsymbol{\Pi} = \langle \boldsymbol{\pi} \rangle = -\langle \boldsymbol{\sigma}\nabla n \rangle \tag{2.13}$$

The unique vectors, operators $\boldsymbol{\pi}$ and the $\boldsymbol{\Pi}$, are nonlocal and they map the nonrandom scalar field $U$ into the random vector fields $\mathbf{v}$ and nonrandom field $\mathbf{V}$. From the condition $\nabla \mathbf{v} = \nabla \mathbf{V} = f$ and Equations (2.12), (2.13) and (2.9) we have the relations: $\nabla \boldsymbol{\Pi} = \nabla \boldsymbol{\pi} = \mathbf{L}$.

Thus we have a closed system for the local averaged fields $U$ and $\mathbf{V}$: Equations (2.13) for the averaged scalar field $U$ and Equations (2.12) for the averaged vector field $\mathbf{V}$. Evidently all operators $m, \mathbf{L}, \boldsymbol{\pi}$ are linear. They do not depend on the function $f$, but in general they are related to the domain $\Omega$.

Now we analyze the partial case when $3-D$ domain $\Omega$ is unbounded and the random field $\boldsymbol{\sigma}(\mathbf{x})$ is stochastically homogeneous. As we emphasized in Shvidler and Karasaki (2008) the local operator $l$ is self-adjoint and therefore, the random Green's function is symmetric by arguments: $g(\mathbf{x}, \mathbf{y}) = g(\mathbf{y}, \mathbf{x})$, as is the real and even mean Green's function $G(\mathbf{x}, \mathbf{y})$. From equation (2.8) we have

$$U(\mathbf{x}) = \int_\Omega G(\mathbf{x} - \mathbf{y}) f(\mathbf{y}) dy^3 \tag{2.14}$$

Applying to (2.14) the generalized Fourier transformation, we have

$$\bar{J}(\mathbf{k})\bar{U}(\mathbf{k}) = \bar{f}(\mathbf{k}), \quad \bar{J}(\mathbf{k}) = \bar{G}^{-1}(\mathbf{k}) \tag{2.15}$$

and in the real $\mathbf{x}$-space the averaged nonlocal equation for $U(\mathbf{x})$ is

$$\int_\Omega J(\mathbf{x}-\mathbf{y})U(\mathbf{y})dy^3 = f(\mathbf{x}) \tag{2.16}$$

Substituting now $f(\mathbf{x})$ from Equation (2.16) into Equation (2.7) we have the relation between the random function $u$ and nonrandom function $U$

$$u(\mathbf{x}) = \int_\Omega\int_\Omega g(\mathbf{x},\mathbf{y})J(\mathbf{y}-\mathbf{z})U(\mathbf{z})dy^3 dz^3 \tag{2.17}$$

Thus, the random operator $n$ from equation (2.10) is

$$n = \int_\Omega\int_\Omega g(\mathbf{x},\mathbf{y})J(\mathbf{y}-\mathbf{z})(\bullet)dy^3 dz^3 \tag{2.18}$$

That is the operator $n$ depends on the variable $\mathbf{x}$ and operates in space of nonrandom function $U(\mathbf{z})$. Applying now the random vector-operator $-\boldsymbol{\sigma}(\mathbf{x})\nabla_x$ to both sides of Equation (2.10), we can find the random local velocity vector $\mathbf{v}(\mathbf{x})$

$$\mathbf{v}(\mathbf{x}) = -\boldsymbol{\sigma}(\mathbf{x})\nabla_x[nU] \tag{2.19}$$

or using the Equations (2.17) and (2.18)

$$\mathbf{v}(\mathbf{x}) = \int_\Omega\int_\Omega \boldsymbol{\gamma}(\mathbf{x},\mathbf{y})J(\mathbf{y}-\mathbf{z})U(\mathbf{z})dy^3 dz^3 \tag{2.20}$$

Here $\boldsymbol{\gamma}(\mathbf{x},\mathbf{y}) = -\boldsymbol{\sigma}(\mathbf{x})\nabla_x g(\mathbf{x},\mathbf{y})$ is random Greens velocity function that satisfied the equation

$$\nabla\boldsymbol{\gamma}(\mathbf{x},\mathbf{y}) = \delta(\mathbf{x}-\mathbf{y}) \tag{2.21}$$

Averaging now Equations (2.20), (2.21) and set $\langle\boldsymbol{\gamma}(\mathbf{x},\mathbf{y})\rangle = \boldsymbol{\Gamma}(\mathbf{x}-\mathbf{y})$ we have

$$\mathbf{V}(\mathbf{x}) = \iint \boldsymbol{\Gamma}(\mathbf{x}-\mathbf{y})J(\mathbf{y}-\mathbf{z})U(\mathbf{z})dy^3 dz^3, \quad \nabla_\mathbf{x}\boldsymbol{\Gamma}(\mathbf{x}-\mathbf{y}) = \delta(\mathbf{x}-\mathbf{y}) \tag{2.22}$$

Using the Fourier transformation for Equations (2.22) and bearing in mind (2.15) we have

$$\bar{\mathbf{V}}(\mathbf{k}) = \bar{\boldsymbol{\Pi}}(\mathbf{k})\bar{U}(\mathbf{k}), \quad \bar{\boldsymbol{\Pi}}(\mathbf{k}) = \bar{\boldsymbol{\Gamma}}(k)\bar{G}^{-1}(\mathbf{k}), \quad 2\pi i\mathbf{k}\bar{\boldsymbol{\Gamma}}(\mathbf{k}) = 1 \tag{2.23}$$

Here the scalar $\bar{G}(\mathbf{k})$ and vector $\bar{\boldsymbol{\Gamma}}(\mathbf{k})$ are transformed Greens mean and Greens mean velocity functions, respectively.

In the real space we have exact equation for mean flow velocity

$$\mathbf{V}(\mathbf{x}) = \int \mathbf{\Pi}(\mathbf{x}-\mathbf{y})U(\mathbf{y})dy^3 \qquad (2.24)$$

Now we present vector $\bar{\mathbf{\Gamma}}(\mathbf{k})$ in the form $\bar{\mathbf{\Gamma}}(\mathbf{k}) = \bar{\mathbf{\Gamma}}^*(\mathbf{k}) + \bar{\mathbf{\Gamma}}_*(\mathbf{k})$, where vector-function $\bar{\mathbf{\Gamma}}^*(\mathbf{k})$ is imaginary and odd in $\mathbf{k}$-space, and $\bar{\mathbf{\Gamma}}_*(\mathbf{k})$ - a real and even vector-function. Inserting this formula in the third Equation (2.23) and comparing the real and imaginary terms, for any $\mathbf{k}$ we have two conditions: $\bar{\mathbf{\Gamma}}_*(\mathbf{k}) = 0$ that is $\bar{\mathbf{\Gamma}}(\mathbf{k}) = \bar{\mathbf{\Gamma}}^*(\mathbf{k})$ or a system of two equations: $2\pi i \mathbf{k}\bar{\mathbf{\Gamma}}_*(\mathbf{k}) = 0$ and $2\pi i \mathbf{k}\bar{\mathbf{\Gamma}}^*(\mathbf{k}) = 1$. From the first equation we have: $\bar{\mathbf{\Gamma}}_*(\mathbf{k}) = 0$ or $\bar{\mathbf{\Gamma}}(\mathbf{k}) = -\bar{\mathbf{\Gamma}}(-\mathbf{k})$, that is for a stochastic homogeneous random field $\bar{\mathbf{\Gamma}}(\mathbf{k})$ is a skew-symmetric imaginary vector. On the contrary, the condition $\bar{\mathbf{\Gamma}}_*(\mathbf{k}) \neq 0$ and is orthogonal to vector $\mathbf{k}$, contradict with skew-symmetry condition for $\bar{\mathbf{\Gamma}}(\mathbf{k})$, and is incorrect.

Now we introduce the real and even symmetric tensor $\bar{\mathbf{B}}(\mathbf{k})$, that satisfies the equation $\bar{\mathbf{\Pi}}(\mathbf{k}) = -\bar{\mathbf{B}}(\mathbf{k})2\pi i \mathbf{k}$, and after inserting $\bar{\mathbf{\Pi}}(\mathbf{k})$ in the first equation (2.23) we have in $\mathbf{x}$-space

$$\mathbf{V}(\mathbf{x}) = -\int \mathbf{B}(\mathbf{x}-\mathbf{y})\nabla U(\mathbf{y})dy^3 \qquad (2.25)$$

It is evident that last equation is equivalent with equation (2.24). This is sufficient for integrating by parts in (2.25).

Arrange $\bar{\mathbf{\Pi}}(\mathbf{k})$ with second and third formulas from (2.23) we have by each $\mathbf{k}$ the relation $4\pi^2 \mathbf{k}\mathbf{B}(\mathbf{k})\mathbf{k} = \bar{\mathbf{G}}^{-1}(\mathbf{k})$, whose left part is in a real positive definite symmetric quadratic form, and therefore, there exists a principal orthogonal coordinate system $\mathbf{k}'$ in which the tensor $\bar{\mathbf{B}}(\mathbf{k}')$ is diagonal. In addition, because the local random field $\boldsymbol{\sigma}(\mathbf{x})$ is stochastically homogeneous, the orientations of the axes of the tensor $\bar{\mathbf{B}}(\mathbf{k}')$ for any $\mathbf{k}'$, including the homogenization limit $\mathbf{k}' \to 0$, are identical. In this case, if $l \neq m$, $\bar{B}_{lm}(\mathbf{k}') = 0$, with the real diagonal components $B_{ll}(\mathbf{k}') = -\bar{\Pi}'_l(\mathbf{k}')/2\pi i k'_l$ (no summation over $l$).

But what if the orientation of the principal coordinate system is unknown a priori? In this case analyzing the asymptotic behavior of the vector $\bar{\mathbf{\Pi}}'(\mathbf{k})$ for small $|\mathbf{k}|$, we can find $\bar{\Pi}'_l(\mathbf{k}) = -\bar{B}_{lm}2\pi i k_m$, where $\bar{B}_{lm} = -\dfrac{1}{2\pi i}\dfrac{\partial}{\partial k_m}\bar{\Pi}'_l(0)$ is a real constant tensor component of effective conductivity, which is symmetric by index. Therefore, we can find its real eigenvalues and orthogonal eigenvectors and after transformation to the new coordinates associated with them, we find for any $\mathbf{k}$ the diagonal tensor

$$\bar{B}_{ll}(\mathbf{k}) = -\bar{\Pi}_l(\mathbf{k})/2\pi i k_l , \quad \bar{B}_{lm}(\mathbf{k}) = 0, \text{ if } l \neq m \text{ and in real space}$$

$$V_l(\mathbf{x}) = -\int B_{ll}(\mathbf{x}-\mathbf{y})\frac{\partial}{\partial y_l}U(\mathbf{y})dy^3 \qquad (2,26)$$

As analyzed by (Shvidler and Karasaki, 2008), in 3-dimensional unbounded stochastically homogeneous spaces it is possible to select only three types of so-called global symmetry: 1) isotropic, for which in any coordinate system $\bar{B}_{ll}(\mathbf{k}) = \bar{B}_{ll}(|\mathbf{k}|)$, 2) orthotropic, for which in corresponding principal axes $\bar{B}_{ll}(\mathbf{k}) = \bar{B}_{ll}(|k_1|,|k_2|,|k_3|)$ are positive and even, and 3) transversal isotropic, for which three positive and even functions $\bar{B}_{ll}(\mathbf{k})$ are invariant relative to the rotation around only one axis of the coordinate system ( for example, $k_3$ ) and reflection on the any planes $k_l = 0$. They have the form

$$\bar{B}_{ll}(\mathbf{k}) = \bar{B}_{ll}\left(\left(k_1^2 + k_2^2\right)^{1/2}, |k_3|\right).$$

Thus as well as in the homogenization limit for the general case in $\mathbf{k}$ and $\mathbf{x}$ three dimensional space, only three types of global symmetry exist: isotropic, orthotropic and transversal isotropic.

**Remark**. It should be noted that all previous results are valid for any two and tree dimensional fields except the unbounded two-dimensional fields for which the Green functions differ at infinity. This case was analyzed earlier ( Shvidler M. and K. Karasaki, 2008**).**

## 3. Non-steady Transient Flow

We consider a stochastic system of equations for pressure $u(\mathbf{x},t)$ and flow-velocity vector $\mathbf{v}(\mathbf{x},t)$ in a single connected $d$-dimensional domain $\Omega$

$$\beta\theta(\mathbf{x})\frac{\partial u(\mathbf{x},t)}{\partial t} + \nabla\mathbf{v}(\mathbf{x},t) = f(\mathbf{x},t) \quad, \quad \mathbf{v}(\mathbf{x},t) = -\boldsymbol{\sigma}(\mathbf{x})\nabla u(\mathbf{x},t) \qquad (3.1)$$

Here a nonrandom $\beta = const$ is coefficient of a volume deformation of a liquid –core system, $\theta(\mathbf{x})$ and $\boldsymbol{\sigma}(\mathbf{x})$ are porosity and conductivity random the elliptic tensor, respectively. The non-random specified density of sources and sinks $f(\mathbf{x},t)$, is an integrable and compactly supported function. The pressure $u(\mathbf{x},t)$ satisfies the initial condition $u(\mathbf{x},0) = 0$, if $\mathbf{x} \in \Omega$, and the boundary condition $u(\mathbf{x},t) = 0$, if $\mathbf{x} \in \partial\Omega$. If domain $\Omega$ is unbounded, at infinity $u(\mathbf{x},t) \to 0$. Combining now Equations (3.1) we have for the displaced pressure $u(\mathbf{x},t)$ a parabolic second order differential equation

$$\beta\theta(\mathbf{x})\frac{\partial u(\mathbf{x},t)}{\partial t} - \nabla\left[\boldsymbol{\sigma}(\mathbf{x})\nabla u(\mathbf{x},t)\right] = f(\mathbf{x},t) \qquad (3.2)$$

Thus we have the initial-boundary problem for the $u(\mathbf{x},t)$. The unique solution of Equation (3.2) with homogeneous boundary and initial conditions exists. For this solution we can note

$$lu = f \quad, \quad l = \beta\theta(\mathbf{x})\frac{\partial}{\partial t} - \nabla[\boldsymbol{\sigma}(\mathbf{x})\nabla] \tag{3.3}$$

The operator $l$ maps the space of the Equation (3.2) random unique solutions $u(\mathbf{x},t)$ in the space of nonrandom function $f(\mathbf{x},t)$ and according to the Green's formula:

$$u(\mathbf{x},t) = \int_0^t \int_\Omega g(\mathbf{x},t;\mathbf{y},\tau) f(y,\tau) dy^d d\tau \tag{3.4}$$

Here the random Green's function $g(\mathbf{x},t;\mathbf{y},\tau)$ satisfies the equations:

$$l g(\mathbf{x},t;\mathbf{y},\tau) = \delta(\mathbf{x}-\mathbf{y})\delta(t-\tau) \quad, \mathbf{x},\mathbf{y} \in \Omega, \tau \leq t \tag{3.5}$$

$$g(\mathbf{x},0;\mathbf{y},\tau) = 0, \quad \mathbf{x},\mathbf{y} \in \Omega \quad and \quad g(\mathbf{x},0;\mathbf{y},0) = 0$$

Now we can write (3.4) the solution of equation (3.2), that also satisfies the uniform conditions

$$u = mf, \quad m = \int_0^t \int_\Omega g(\mathbf{x},t;\mathbf{y},\tau)(\cdot) dy^d d\tau \tag{3.6}$$

As in Section 2, it is evident that $lm = \mathbf{I}^{(f)}$ and $ml = \mathbf{I}^{(u)}$, where $\mathbf{I}^{(f)}$ and $\mathbf{I}^{(u)}$ are identity mappings in $f$ and $u$ functional spaces, respectively. That is the random operators $l$ and $m$ are non-commutative. So, averaging Equations (3.6) of the ensembles of realizations $\boldsymbol{\sigma}$ and $\theta$, we have for the mean pressure $U = \langle u \rangle$

$$U = \langle m \rangle f \quad, \quad \langle m \rangle = \int_0^t \int_\Omega G(\mathbf{x},t;\mathbf{y}\tau)(\bullet) dy^d d\tau \tag{3.7}$$

Multiplaing the Equation (3.7) by the unique operator $\mathbf{L} = \langle m \rangle^{-1}$, which in the general case is nonlocal, we derive the averaged functional equation for the mean field $U$

$$\mathbf{L}U = f \tag{3.8}$$

with the relevant boundary and initial conditions

$$U_{\partial\Omega} = 0 \quad, U_{t=0} = 0 \tag{3.9}$$

By comparing Equations (3.7) and (3.6) we obtain the relationship between the spaces of random field $u(\mathbf{x},t)$ and the nonrandom field $U$

$$u(\mathbf{x},t) = nU , \quad n = m\langle m \rangle^{-1}, \langle n \rangle = \mathbf{I}^{(U)} \tag{3.10}$$

And the relationship between the random velocity vector $\mathbf{v}(\mathbf{x},t)$ and nonrandom field $U$

$$\mathbf{v}(\mathbf{x},t) = \boldsymbol{\pi} U , \quad \boldsymbol{\pi} = -\boldsymbol{\sigma}(\mathbf{x})\nabla_x n \tag{3.11}$$

As above in Section 2, we are bearing in mind that except at the homogenization limit the local nonrandom function $U$ in Equation (3.11) in contrast of operator $n$ is independent of variable **x**. Averaging as above Equations (3.11) we can find for the mean velocity vector $\mathbf{V}(\mathbf{x},t) = \langle \mathbf{v}(\mathbf{x},t) \rangle$

$$\mathbf{V}(\mathbf{x},t) = \boldsymbol{\Pi} U , \quad \boldsymbol{\Pi} = \langle \boldsymbol{\pi} \rangle = -\langle \boldsymbol{\sigma}(\mathbf{x})\nabla_x n \rangle \tag{3.12}$$

Combining formulas (3.1) and (3.10) we have

$$\langle \theta(\mathbf{x}) u(\mathbf{x},t) \rangle = \theta_* U , \tag{3.13}$$

$$\theta_* = \langle \theta(\mathbf{x}) n \rangle \tag{3.14}$$

It is evident that the nonrandom operator $\theta_*$ is a nonlocal functional. We will call it - the effective transient flow porosity.

Returning now to Equation (3.1) and averaging it, using in addition (3.12), (3.13) and (3.14) we have the functional equation for the mean function $U$, which satisfies the initial and boundary conditions ,

$$\mathbf{L} U = f , \quad \mathbf{L} = \beta \frac{\partial}{\partial t} \theta_* + \nabla \boldsymbol{\Pi} \tag{3.15}$$

$$U(\mathbf{x},0) = 0, U_{\partial\Omega}(\mathbf{x},t) = 0$$

Clearly, Equations (3.15) generalize the similar conditions for steady-state flow as those noted in Section 2. Now we consider partly case when transient flow in the 3-D unbounded domain $\Omega$ and assume that the random conductivity $\boldsymbol{\sigma}(\mathbf{x})$ and porosity $\theta(\mathbf{x})$ fields are stochastically homogeneous in $\Omega$. We set that the local equations (3.1)-(3.3) are valid for any $t > 0$ in $\Omega$ and at $|\mathbf{x}| \to \infty, u(\mathbf{x},t) \to 0$. The initial condition is $u(\mathbf{x},0) = 0.$

Using the random Green's functions $g(\mathbf{x},t;\mathbf{y},\tau), \boldsymbol{\gamma}(\mathbf{x},t;\mathbf{y},\tau) = -\boldsymbol{\sigma}(\mathbf{x})\nabla_x g(\mathbf{x},t;\mathbf{y},\tau)$ and function $p(\mathbf{x},t;\mathbf{y},\tau) = \theta(\mathbf{x})g(\mathbf{x},t;\mathbf{y},\tau)$, considering that $\boldsymbol{\sigma}(\mathbf{x})$ and $\theta(\mathbf{x})$ are stochastically homogeneous, we have $\langle p \rangle = P(\mathbf{x}-\mathbf{y},t-\tau), \langle g \rangle = G(\mathbf{x}-\mathbf{y},t-\tau), \langle \boldsymbol{\gamma} \rangle = \boldsymbol{\Gamma}(\mathbf{x}-\mathbf{y},t-\tau)$.

Applying now for functions $\langle p \rangle, \langle g \rangle$ and $\langle \boldsymbol{\gamma} \rangle$ - the Fourier-Laplace transforms $\bar{T}_{FL}$ we have in $\mathbf{k}$ and $\mu$ spaces functions $\bar{P}(\mathbf{k},\mu), \bar{G}(\mathbf{k},\mu)$, vector $\bar{\boldsymbol{\Gamma}}(\mathbf{k},\mu)$ and the vector $\bar{\boldsymbol{\Pi}}(\mathbf{k},\mu)$ and scalar functions $\bar{S}(\mathbf{k},\mu)$ respectively

$$\bar{\boldsymbol{\Pi}}(\mathbf{k},\mu) = \bar{\boldsymbol{\Gamma}}(\mathbf{k},\mu)\bar{G}^{-1}(\mathbf{k},\mu),\ \bar{S}(\mathbf{k},\mu) = \bar{P}(\mathbf{k},\mu)\bar{G}^{-1}(\mathbf{k},\mu) \qquad (3.16)$$

It is straightforward to show that from equation (3.5) we have in Fourier-Laplace space:

$$\mu\beta\bar{S}(\mathbf{k},\mu)\bar{G}(\mathbf{k},\mu) + 2\pi i\mathbf{k}\bar{\boldsymbol{\Pi}}(\mathbf{k},\mu)\bar{G}(\mathbf{k},\mu) = 1 \qquad (3.17)$$

Multipying equation (3.17) by $\bar{f}(\mathbf{k},\mu) = \bar{T}_{FL}f(\mathbf{x},t)$ and using the formulas
$$\bar{U}(\mathbf{k},\mu) = \bar{G}(\mathbf{k},\mu)\bar{f}(\mathbf{k},\mu),\ \bar{\mathbf{V}}(\mathbf{k},\mu) = \bar{\boldsymbol{\Pi}}(\mathbf{k},\mu)\bar{U}(\mathbf{k},\mu),\ \bar{\mathsf{M}}(\mathbf{k},\mu) = \bar{S}(\mathbf{k},\mu)\bar{U}(\mathbf{k},\mu) \qquad (3.18)$$
we obtain the averaged equation

$$\mu\beta\bar{\mathsf{M}}(\mathbf{k},\mu) + 2\pi i\mathbf{k}\bar{\mathbf{V}}(\mathbf{k},\mu) = \bar{f}(\mathbf{k},\mu) \qquad (3.19)$$

Clearly, for any $\mathbf{x}$ and $(t-\tau)$ in stochastically homogeneous fields $\boldsymbol{\sigma}(\mathbf{x})$ and $\theta(\mathbf{x})$ the kernel-vector $\boldsymbol{\Pi}(\mathbf{x}-\mathbf{y},t-\tau)$ is an odd vector function in $\mathbf{x}-\mathbf{y}$ space. On the other hand the transformed moment $\bar{P}(\mathbf{k},\mu)$ and mean Green's function $\bar{G}(\mathbf{k},\mu)$ are even functions in $\mathbf{k}$-space and for this reason the functions $\bar{S}(\mathbf{k},\mu)$ and $S(\mathbf{x}-\mathbf{y},t-\tau)$ are even in the same way. Taking this into account the averaged nonlocal system for scalar and vector mean function $U$ and $\mathbf{V}$ is

$$\beta\frac{\partial \mathsf{M}(\mathbf{x},t)}{\partial t} + \nabla \mathbf{V}(\mathbf{x},t) = f(\mathbf{x},t) \qquad (3.20)$$

$$\mathsf{M}(\mathbf{x},t) = \int_0^t\!\!\int S(\mathbf{x}-\mathbf{y},t-\tau)U(\mathbf{y},\tau)dy^3 d\tau \qquad (3.21)$$

$$\mathbf{V}(\mathbf{x},t) = \int_0^t\!\!\int \boldsymbol{\Pi}(\mathbf{x}-\mathbf{y},t-\tau)U(\mathbf{y},\tau)dy^3 d\tau \qquad (3.22)$$

$$U(\mathbf{x},0) = U(\infty,t) = 0 \qquad (3.23)$$

Now we continue our analysis, assuming that the fluctuations of porosity function are sufficiently small and porosity $\theta(\mathbf{x})$ is a non-random constant $\theta$.

Applying now the generalized Fourier transform to averaged equation (3.15) and using transformed functions $\overline{G}$, $\overline{\Gamma}$ we have the equation

$$\beta\theta\frac{\partial \overline{G}(\mathbf{k},t-\tau)}{\partial t}+2\pi i\mathbf{k}\overline{\Gamma}(\mathbf{k},t-\tau)=I(\mathbf{k})\delta(t-\tau) \qquad (3.24)$$

Here $I(\mathbf{k})$ is unit in **k**–space. Representing now the vector $\overline{\Gamma}(\mathbf{k},t-\tau)=\overline{\Gamma}_*(\mathbf{k},t-\tau)+\overline{\Gamma}^*(\mathbf{k},t-\tau)$, where $\overline{\Gamma}_*$ is real and even vector-function in **k**–space, but $\overline{\Gamma}^*$ is odd and imaginary on **k** vector–function, the scalar real and even on **k** function $\overline{G}(\mathbf{k},t-\tau)$, and separating the real and imaginary parts in equation (3.24), we have equation $2\pi i\mathbf{k}\overline{\Gamma}_*(\mathbf{k},t-\tau)=0$. Similar to the case of steady flow it is obvious that there is only one solution $\overline{\Gamma}_*(\mathbf{k},t-\tau)=0$, equally matched that $\overline{\Gamma}(\mathbf{k},t-\tau)$ is an odd and imaginary vector in **k**–space, or in other words, is skew-symmetric and therefore, can be presented in the form:

$$\overline{\Gamma}(\mathbf{k},t-\tau)=-\overline{\mathbf{B}}(\mathbf{k},t-\tau)2\pi i\mathbf{k}\overline{G}(\mathbf{k},t-\tau) \qquad (3.25)$$

Here symmetric by index tensor $\overline{\mathbf{B}}(\mathbf{k},t-\tau)$ is a real and even in **k**–space. The function $\overline{G}(\mathbf{k},t-\tau)$ is real and even as well.

Applying now for the equation (3.24) the Laplace transform $\psi(\mathbf{k},\mu)=\int_0^\infty \psi(\mathbf{k},t)e^{-\mu t}dt$, multiplying the result by $\overline{f}(\mathbf{k},\mu)=\overline{T}_{FL}f(\mathbf{x},t)$ and using the formulas (3.18) by $\theta(\mathbf{x})=\theta=const$

$$\overline{V}(\mathbf{k},\mu)=\overline{\Gamma}(\mathbf{k},\mu)\overline{f}(\mathbf{k},\mu),\ \overline{U}(\mathbf{k},\mu)=\overline{G}(\mathbf{k},\mu)\overline{f}(\mathbf{k},\mu) \qquad (3.26)$$

$$\overline{\mathsf{M}}(\mathbf{k},\mu)=\theta\overline{U}(\mathbf{k},\mu),\ \overline{\Gamma}(\mathbf{k},\mu)=-\overline{\mathbf{B}}(\mathbf{k},\mu)2\pi i\mathbf{k}\overline{G}(\mathbf{k},\mu) \qquad (3.27)$$

we obtain the averaged equations in Fourier-Laplace space:

$$\mu\beta\theta\overline{U}(\mathbf{k},\mu)+2\pi i\mathbf{k}\overline{V}(\mathbf{k},\mu)=\overline{f}(\mathbf{k},\mu) \qquad (3.28)$$

$$\overline{V}(\mathbf{k},\mu)=-\overline{\mathbf{B}}(\mathbf{k},\mu)2\pi i\mathbf{k}\overline{U}(\mathbf{k},\mu) \qquad (3.29)$$

Now let at $t \to \infty$ the function $f(\mathbf{x},t)$ have a limit not dependent on time and function $\overline{f}(\mathbf{k},0)$ be finite as well. It is evident that in this case the system (3.27) and (3.28) respectively reduce to equations

$$2\pi i k_l \overline{V}_l(\mathbf{k},0) = \overline{f}(\mathbf{k},0) \qquad (3.30)$$

$$\overline{V}_l(\mathbf{k},0) = -\overline{B}_{lm}(\mathbf{k},0) 2\pi i k_m \overline{U}(\mathbf{k},0), \qquad (3.31)$$

which describe the stabilized steady flow process in k–space.

As demonstrated in Section 2, if the field $\boldsymbol{\sigma}(\mathbf{x})$ is stochastically homogeneous in $\mathbf{x}$–space, we can transform the coordinate system $\mathbf{k}$ to $\mathbf{k}'$ in which the tensor $\overline{\mathbf{B}}(\mathbf{k}',0)$ for any $\mathbf{k}'$ is diagonal. Now we set this phenomenon valid for any $\mu$ and unique tensor $\overline{B}'_{lm}(\mathbf{k}',\mu)$ be diagonal in the same coordinate system $\mathbf{k}'$, in which the tensor $\overline{B}'_{lm}(\mathbf{k}',0)$ is diagonal. In this coordinate system we can find the components $\overline{B}_{lm}(\mathbf{k}',\mu)$

$$\overline{B}'_{ll}(\mathbf{k}',\mu) = -\overline{\Gamma}'_l(\mathbf{k}',\mu) / 2\pi i k'_l \overline{G}'(\mathbf{k}',\mu) \ , \ \ \overline{B}'_{lm}(\mathbf{k}',\mu) = 0 \ \ \text{if} \ \ l \neq m \qquad (3.32)$$

Clearly, for any $\mathbf{x}$ and $t$ in stochastically homogeneous field $\boldsymbol{\sigma}(\mathbf{x})$ and $\theta = const$, dropping the upper primes, we have the averaged nonlocal system for scalar and vector mean functions $U(\mathbf{x},t)$ and $\mathbf{V}(\mathbf{x},t)$

$$\beta\theta \frac{\partial U(\mathbf{x},t)}{\partial t} + \nabla \mathbf{V}(\mathbf{x},t) = f(\mathbf{x},t) \qquad (3.33)$$

$$\mathbf{V}(\mathbf{x},t) = -\int_0^t \int B(\mathbf{x}-\mathbf{y},t-\tau) \nabla U(\mathbf{y},\tau) dy^3 d\tau \qquad (3.34)$$

$$U(\mathbf{x},0) = U(\infty,t) = 0 \qquad (3.35)$$

It should be noted that homothetic to the case of steady flow we can analyze three types of global symmetry of the tensor $\mathbf{B}(\mathbf{x}-\mathbf{y},t-\tau)$: isotropic, transversal isotropic and orthotropic.

**4. Non-reactive Stochastic Solute Transport**

Here we examine the stochastic transport of nonreactive solute in random porous media and therefore the flow velocity and the solute concentration are random fields as well. We consider both processes (flow and transport) jointly. We assume that at the initial moment the solute is absent in the field, but that later, solute is input or output through distributed sources or sinks in the field.

The method used to input solute in the field may vary. It may be injected with some volumes of carrier liquid. If these volumes are relatively significant they can reasonably be included in the flow balance equation. If the solute is output from the system together with liquid, we must account for it, because the given or prescribed density of liquid sinks and solute sinks are mutually dependent.

Here, we analyzed in details the case when solute is injected in space. For example, it is similar by creating some "plumes'' and studying their evolution.

Now we study flow and transport in the three-dimensional unbounded domain $\Omega$. The solute local transport equations are as follows:

$$\theta(\mathbf{x})\frac{\partial c(\mathbf{x},t)}{\partial t}+\nabla \mathbf{q}(\mathbf{x},t)=\varphi(\mathbf{x},t)\,,\quad \mathbf{q}(\mathbf{x},t)=\mathbf{v}(\mathbf{x},t)c(\mathbf{x},t) \tag{4.1}$$

The concentration of solute $c(\mathbf{x},t)$ is absent in the field at $t=0$ and vanishes at infinity for any time.

To close the system (4.1) we supplement the subsystem of equations (3.1) which describes independent velocity $\mathbf{v}(\mathbf{x},t)$. We assume that the conductivity tensor $\boldsymbol{\sigma}(\mathbf{x})$ and porosity function $\theta(\mathbf{x})$ are random functions. Therefore the flow-velocity $\mathbf{v}(\mathbf{x},t)$, the solute flux $\mathbf{q}(\mathbf{x},t)$, as well function $c(\mathbf{x},t)$ are random. We assume that the function $\varphi(\mathbf{x},t)$ is the nonrandom solute source density. As we note later for this case the density of sinks $f(\mathbf{x},t)$ and $\varphi(\mathbf{x},t)$ must be independent.

Let us assume that the initial value of pressure, $u(\mathbf{x},0)=0$. Combining Equations (4.1) and (3.11) we can write for $c(\mathbf{x},t)$ the linear equation and the initial and boundary conditions as follows:

$$\hat{l}c(\mathbf{x},t)=\varphi(\mathbf{x},t)\,,\quad \hat{l}=\theta(\mathbf{x})\frac{\partial}{\partial t}+\nabla \pi U \tag{4.2}$$

$$c(\mathbf{x},0)=0\,,\, c(\mathbf{x},t)=0 \quad \text{if } x\to\infty \tag{4.3}$$

Introducing now the Green's function $\hat{g}$ that satisfies the equation $\hat{l}\hat{g}=\delta(\mathbf{x}-\mathbf{y})\delta(t-\tau)$ with the uniform initial and boundary conditions we have

$$\hat{c}=\hat{m}\varphi\,,\quad \hat{m}=\int_0^t\!\!\int_\Omega \hat{g}(\mathbf{x},\mathbf{y};t,\tau)\ dy^3 d\tau \tag{4.4}$$

and after averaging the last equation over random fields $\boldsymbol{\sigma}$ and $\theta$ and designating $\hat{C}=\langle\hat{c}\rangle$, we write

$$\hat{C}=\langle\hat{m}\rangle\varphi \tag{4.5}$$

Multiplying the Equation (4.5) by the unique operator $\langle \hat{m} \rangle^{-1}$ leads to the basic equation for mean concentration

$$\hat{\mathbf{L}}\hat{C} = \varphi, \quad \hat{\mathbf{L}} = \langle \hat{m} \rangle^{-1} \tag{4.6}$$

The unique operator $\hat{\mathbf{L}}$ we shall call as effective.

From Equations (4.5) and (4.4) we find the relationship between the random and mean concentration

$$\hat{c} = \hat{n}\hat{C}, \quad \hat{n} = \hat{m}\langle \hat{m} \rangle^{-1} \tag{4.7}$$

The scalar random operator $\hat{n}$ for almost all realization of the random fields $\boldsymbol{\sigma}(\mathbf{x})$ and $\theta(\mathbf{x})$ maps the uniform nonrandom function space $\hat{C}$ into relevant realizations of random functions space $\hat{c}$. Evidently we have $\langle \hat{n} \rangle = \mathbf{I}^{(\hat{C})}$ - identity mapping in the space of nonrandom functions $\hat{C}$.

Using these results and the relation (4.1), we can find the mean flux $\hat{Q}$

$$\hat{Q} = \mathbf{W}\hat{C} \quad \mathbf{W} = \langle \boldsymbol{\pi}\hat{U}\hat{n} \rangle \tag{4.8}$$

The vector $\mathbf{W}$ can be called as the effective transport velocity, which reflects the convective mechanisms of transport because $\mathbf{W}$ is linked with $\hat{n}$.

Similarly, we obtain

$$\langle \theta \hat{c} \rangle = \langle \theta \hat{n}\hat{C} \rangle = \langle \theta \hat{n} \rangle \hat{C} = \theta^* \hat{C} \tag{4.9}$$

The nonrandom functional $\theta^*$ can be called as the effective transport porosity. It is evident that both effective functionals $\theta^*$ and $\theta_*$ from Section 3, unlike the porous media function $\theta(\mathbf{x})$, also reflect appropriately different processes: transport and transient flow.

Averaging Equation (4.1) and using the formulas (4.8) and (4.9) we can write the equation and initial and boundary conditions for the mean concentration $C(\mathbf{x},t)$:

$$\frac{\partial}{\partial t}\left[ \theta^*(\mathbf{x},t)C(\mathbf{x},t) \right] + \nabla\left[ \mathbf{W}(x,t)C(\mathbf{x},t) \right] = \varphi(\mathbf{x},t) \tag{4.10}$$

$$C(\mathbf{x},0) = 0 \text{ , by } \mathbf{x} \in \Omega \text{ and } C(\mathbf{x},t) = 0 \text{ if } x \to \infty \tag{4.11}$$

From Equations (4.10) we can write the semi-explicit form of the effective operator $\hat{\mathbf{L}}_*$, which includes the effective operators $\theta^*$ and $\mathbf{W}$.

$$\hat{\mathbf{L}}_* = \frac{\partial}{\partial t}\theta^*(\mathbf{x},t) + \nabla \mathbf{W}(\mathbf{x},t) \tag{4.12}$$

## 5. Summary and Conclusions

We have established the general form for the *exactly* averaged system of basic equations for steady flow, non-steady transient flow and solute transport with sources and sinks. We examined the validity of the averaged descriptions and the generalized law for some non-local models. The approach that developed in the present paper does not require assuming the existence of any small parameters, for example, small scales of heterogeneity or small perturbation of conductivity field. More specifically, the following points are pertinent:

1. We analyzed stochastic non-homogeneous systems and showed the possibility of averaged description for the general common case in a finite domain and arrived at the description of flow for bounded and homogeneous fields.

2. Found the relationship between the local field and the mean filed in the form of $u = nU$.

3. Analyzed various geometries of stochastically homogeneous systems in different scales using symmetry at micro and macro scales.

4. The analysis was extended to non-steady problems in finite and infinite time. The asymptotic behavior of non-steady flow in unbounded, stochastically homogeneous field was obtained.

5. We briefly discussed the average transport properties using the same approach.

## Acknowledgments

The authors would like to thank Dr. D. Silin (Shell International Exploration and Production Inc.) for constructive comments. This work was conducted under the US Department of Energy Contract No. DE-AC02-05CH11231.